\documentclass{jfm}
\usepackage{graphicx}
\usepackage{epstopdf, epsfig}
\usepackage{xcolor}
\usepackage{psfrag}
\usepackage{tikz}
\usepackage{amsmath}
\newcommand\encircle[1]{%
  \tikz[baseline={([yshift=-1pt] X.base)}] 
    \node (X) [draw, shape=circle, inner sep=0] {\strut #1};}

\shorttitle{Inflated balls impacts}
\shortauthor{L. Tadrist and B. Darbois Texier}

\title{Open questions on the impact of an inflated ball}

\author{Lo\" ic Tadrist\aff{1}
  \corresp{\email{loic.tadrist@ulg.ac.be}} and Baptiste Darbois Texier\aff{2}}

\affiliation{\aff{1} Microfluidics Lab, Department of Mechanical and Aerospace Engineering, Building B52, University of Li\`ege, 4000 Li\`ege, Belgium \aff{2} GRASP, Department of Physics, Building  B5, University of Li\`ege, 4000 Li\`ege, Belgium}

\begin{document}
\maketitle

\begin{abstract}
The behaviour of sports balls during impact defines some special features of each sport. The velocity of the game, the accuracy of passes or shots, the control of the ball direction after impact, the risks of injury, are all set by the impact mechanics of the ball. For inflated sports balls, those characteristics are finely tuned by the ball inner pressure. As a consequence, inflation pressures are regulated for sports played with inflated balls. Despite a good understanding of ball elasticity, the source of energy dissipation for inflated balls remains controversial. We first give a clear view of non-dissipative impact mechanics. Second we review, analyse and estimate the different sources of energy dissipation of the multi-physics phenomena that occur during the impact. Finally, we propose several experiments to decide between gas compression, shell visco-elastic dissipation, solid friction, sound emission or shell vibrations as the major source of energy dissipation.
\end{abstract}


\section{Introduction}

Inflated objects exhibit convenient characteristics. It is worth noting that when deflated, their volume is reduced that allows easy storage (inflatable bed, inflatable helmet or car air-bag for example). The main advantage of inflated bodies is the controllability of mechanical properties by the fine-tuning of inflation pressure. Those objects are much used in sports: almost all sports balls are inflated (football, rugby, basketball, handball, etc.)

In 2015, a huge scandal, the \textit{Deflategate}, broke out in American Football \citep{hassett2015football}. During the 2015 conference final against Indianapolis Colts, New England Patriot's team was suspected to use under-inflated balls (under the pressure prescribed by the rules) to advantage ball transmissions between their quarterback and the receivers. Indeed, under-inflated balls are easier to grab and to catch. This is possible only because in American Football, the rules indicate that each team may use their own balls to play. This cheat, in a professional sport, in which huge amounts money are invested, points out the crucial role of pressure inflation of sports balls.

The sports balls are all submitted to specific game rules. For every sport, the rules of the game strictly bound the inflation pressure. Why is it so?
This question is not straight forward, since sports rules were built by usage. Sports rules followed constant evolution up to converge on the today's prescribed inflation pressure. An hypothesis for the convergence value for inflation pressure could be a fun/risk trade-off. One may think that in the one hand, the sport might be fast to be entertaining which requires large inflation pressure. In the other hand, if the ball is too much inflated it induces injuries after one single or repetitive impacts, which may prevent largest inflation pressures. However, to answer this hypothesis, one has to understand first how does the ball behaves during an impact: what is the impact mechanics of an inflated ball?

The goal of this paper is to investigate the mechanics of inflated balls. More precisely, how does impact characteristics depend onto inflation pressure, impact velocities and shell properties?

The problem of sports ball impacts involves lot's of physics. This is a nice \textit{fluid-structure interaction}, ball shell and inner gas are strongly coupled. During the impact, shell is deformed and gas compressed involving \textit{solid mechanics}, \textit{fluid mechanics} and \textit{thermodynamics}. One may have also noticed that, during the impact, the ball emits sound by \textit{acoustics} vibration of the surrounding air.  Seen from far behind, the impact of an inflated ball is already a complex multi-physics problem. 

A short literature review, shows that no scientific consensus exists upon inflated ball mechanics. It is almost well understood that ball elastic properties come from the gas compression inside the ball during the impact for the common inflated balls \citep{cross1999bounce,goodwill2001spring,stronge2007oblique} or partly from the shell and the gas (squash ball, see \citet{lewis2011dynamic}), or uniquely from the shell (tennis-table ball, see \citet{cross2014impact}). However, the source of dissipation is still unclear and several authors have given various explanations to stand for it such as momentum flux force dissipation \citep{stronge2007oblique}, vibrations of the shell \citep{cross2014impact}, dissipation in the gas \citep{georgallas2015coefficient} and solid friction \citep{pauchard1998contact}. No consensus arose to answer this question, especially no model gives the dependency of ball dissipation coefficient (or similarly coefficient of restitution) on inflation pressure.

In this paper, we first recall the mechanics of the impact of an inflated ball without energy loss. The second part of the paper is devoted to the analysis of energy dissipation sources and give when it is possible their expected order of magnitude. Finally the paper discusses the momentum flux force dissipation proposed by \citet{stronge2007oblique} and proposes experiments to find the dominant source of energy dissipation.

\begin{table}
\begin{tabular}{lll}
Symbol & Parameter & Football ball typical value\\
\hline
$R_0$& Radius of the ball & 10.8-11.1 cm\\
$V$, $V_0$ & Volume of the ball, resp. initial volume of the ball& 5.58 L\\
$m$& Mass of the ball& 410-450 g\\
$m_\mathrm{gas}$& mass of gas inside the ball& $\sim$ 13 g \\
$A$&Area of contact between ball and ground&  $\sim$ 10 cm$^2$ \\
$x_g$&Ball center of mass& -\\
$F$& Force exerted by the ball at impact& - \\
$P$, $P_0$& Inner pressure of the ball, resp. initial inner pressure& $\sim$ 1.8 bar\\
$T$, $T_0$& Temperature inside the ball, resp. initial temperature& 293.15 K \\
$\mathcal{E}_d$ & Dissipated energy& 146 J\\
$U$, $U_0$ & Velocity of the ball, resp. initial ball velocity& 51 m/s\\
$\dot{•}$ & denotes devivative towards time, $t$&\\
$'$&denotes derivatives towards $x$&\\
\end{tabular}
\caption{Parameter and symbols used in the paper.} \label{tab:parameters}
\end{table}

\section{Mechanics of non-dissipative impact}\label{sec:non_dissipative}
As stated in the introduction, the mechanics of ball is a complex fluid structure interaction. In order to simplify the modelling, we propose a convenient kinematics of impact of an inflated ball. This kinematics relies on experimental observations.
\subsection{Kinematics of inflated ball impacts}
\subsubsection{Observations of inflated ball impacts}
Real inflated sports balls from different sports (volleyball, football, basketball, handball, waterpolo and futsal) have been chosen for experiments. Those balls were inflated at pressures prescribed by the game. The balls were launched at velocities ranging from 0 to 10 m/s onto a rigid marble (in real sports conditions velocities may range from 0 to 50 m/s). Impacts were recorded with a high speed camera at 4000 frames per second. Special care has been brought for correct illumination of the scene thanks to high power LEDs.

Typical deformation of the ball is shown Fig. \ref{fig:kinematics}. It is observed that the ball remains a spherical cap and no increase in radius has been noticed during the impact. Deviation from the shape of a spherical cap of constant radius has been observed for large indentation, $x>2R_0/3$. In the range of very large indentation, we also observed the formation or ripples on the ball membrane.

Contact times, $t_\mathrm{contact}$, and coefficients of restitution, $e=U_\mathrm{out}/U_0$, were systematically measured with a \textit{matlab 2015b} code based on spherical shape recognition. During the experiments, almost no vibrations have been observed. In the limit of small indentation, we observed that contact times and coefficients of restitution do not depend significantly on the impact velocity, $U_0$. Results are gathered in Table \ref{table:ball_properties}.

\begin{table}
\begin{center}
\begin{tabular}{llrrrrrccrccc}
   && \hphantom{}& \multicolumn{1}{c}{$2R_0$\footnotemark[1]}  & \multicolumn{1}{c}{$m$\footnotemark[1]}  & \multicolumn{1}{c}{$\Delta P$\footnotemark[1]}  &\hphantom{}& $t_\mathrm{contact}$ & $e$  &\hphantom{}& $U_{\mathrm{max}}$ & $\mathcal{E}_{\mathrm{kin}}$  & $\mathcal{E}_{d}$ \\
&&&\multicolumn{1}{c}{(cm)}&\multicolumn{1}{c}{(g)}&\multicolumn{1}{c}{(bar)}&&{(ms)} &-&&{(m/s)}&{(J)}&{(J)}\\
\hline
 \multicolumn{2}{l}{Inflated balls}    &&&&&&&&&&&\\
&\textit{Volley} &&$21\pm 0.3$ & $270\pm 10$ & 0.294 -- 0.318&&  9  & 0.82 && 37 & 185 & 61\\
&\textit{Handball} && $18.8\pm 0.3$ &$ 450\pm 25$ & NR (0.4 -- 0.5) &&  11   & 0.71 && 27 & 164  & 81\\
&\textit{Futsal} && $20.05\pm 0.3$ & $420\pm 20$ & 0.6 -- 0.9  ($^*$)&&  11 & 0.50 &&  42 & 370 & 278\\
&\textit{Water-polo} && $22.1\pm 0.5$  & $425\pm 25$  & 0.55 -- 0.62&& 10   &  0.68  && 24 & 122  & 66\\
&\textit{Basket} &&$24.3\pm 0.5$ & $608\pm 42$ & ($^*$)&&  11  & 0.78 && 16 & 78 & 30.5\\
&\textit{Football} && $21.95\pm 0.3$ & $430\pm 20$ & 0.6 -- 1.1&& 8 & 0.79 && 51 & 559 & 210\\
\hline
\multicolumn{2}{l}{Hollow balls} &&&&&&&&&&\\
&\textit{Squash} &&$4.0\pm 0.05$ & $24\pm 1$ & ($^*$)&&  4  & 0.37 && 78 & 73 & 63\\
&\textit{Table tennis} && 4.0 & 2.7 & NR &&  5  & 0.80 && 32 & 1.4 & 0.5\\
&\textit{Tennis} && $6.7\pm 0.15$ & $57.7\pm 1.7$ & 0 -- 1.0 ($^*$)&& 4.5   & 0.70 && 73 & 154 & 78\\
\end{tabular}
\caption{Properties of various sports balls for impacts with a rigid surface. Mean contact time ($t_\mathrm{contact}$) and mean coefficient of restitution ($e=U_\mathrm{out}/U_0$) have been measured experimentally. Maximal speed obtained in game conditions ($U_\mathrm{max}$), maximal kinetic energy in game conditions ($\mathcal{E}_\mathrm{kin}$) and dissipated energy in game conditions for one impact ($\mathcal{E}_{d}= m U_\mathrm{max}^2 (1-e^2) /2$) have been estimated. NR: Not regulated. (*) Coefficient of Restitution is regulated by the rules of the sport. Basketball: $e_\mathrm{rules}=0.816$ -- 0.882, Futsal: $e_\mathrm{rules}=0.5$ -- 0.57, Squash: $e_\mathrm{rules}=0.346$, Tennis: $e_\mathrm{rules}=0.729$ -- 0.76}
\label{table:ball_properties}
\end{center}
\end{table}

\footnotetext[1]{Data from official rules. \textbf{Volleyball:} Official Volleyball rules 2017-2020, FIVB. \textbf{Handball:} IHF ball regulations 2006, IHF. \textbf{Futsal:} Futsal laws of the game 2014-2015, FIFA. \textbf{Waterpolo:} Waterpolo rules 2015-2017, FINA. \textbf{Basketball:} Official  basketball rules 2012, Basketball equipment, FIBA. \textbf{Football:} Laws of the game 2011-2012, FIFA. \textbf{Squash:} Specification for squash balls 2015, World Squash. \textbf{Table-tennis:} Handbook of the international Table-tennis federation 2016, ITTF. \textbf{Tennis:} ITF approved tennis balls, classified surfaces \& recognised courts, a guide to products \& test methods 2017, ITF.}
   
\subsubsection{Geometrical relationships}
Following the observations above, we propose to describe the shape of the ball during the impact as an indented spherical cap. In the meanwhile, the part of the inflated ball in contact with the ground is flattened. This kinematics is similar to the one adopted by \citet{stronge2007oblique}. The volume, $V$, of a spherical cap and the flat area of contact with the ground, $A$, of a sphere of radius $R_0$ indented by a length $x$ reads,  
\begin{equation}
V(x)=\frac{\pi}{3}\left(4\,R_0^3+ x^3-3\,R_0\,x^2\right)\qquad\mathrm{and}\qquad A(x) = 2\pi\,R_0\, x \,\left(1-\frac{x}{2R_0}\right)\, .
\end{equation}

\begin{figure}
\begin{center}
\begin{psfrags}
\psfrag{x}[c][c]{\textcolor{white}{$x$}}\psfrag{0}[c][c]{0}\psfrag{ex}[c][c]{$\vec{e}_x$}\psfrag{x=xg}[c][c]{\textcolor{white}{$x_0=x_g$}}\psfrag{o}[c][c]{$x_0$}\psfrag{xg}[c][c]{$x_g$}\psfrag{v0}[c][c]{\textcolor{white}{$U_0$}}
\includegraphics[width=\textwidth]{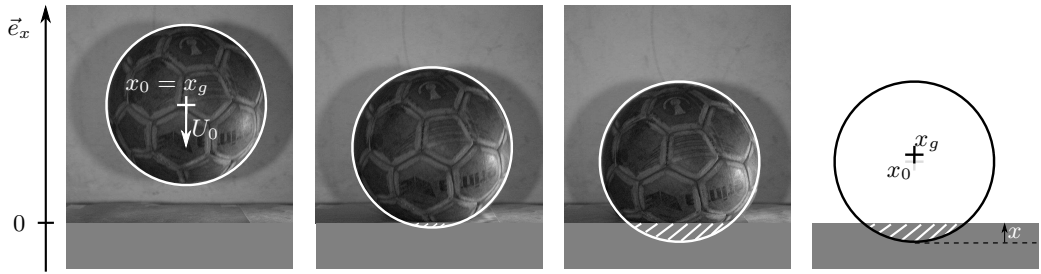}
\end{psfrags}
\caption{Kinematics of ball deformation during the impact. The ball deforms under inertial forces but conserves its spherical shape. No increase of radius has been measured during the impact. Deviation from spherical shapes were observed only for largely squashed balls.}\label{fig:kinematics}
\end{center}
\end{figure}

The deformation of the ball moves its center of mass which is no longer located at the center of the spherical shape, Fig. \ref{fig:kinematics}. The height of the center of mass of the ball can be computed with the hypothesis that all the mass of the ball is constituted by the spherical shell with homogeneous mass distribution. One computes, 
\begin{equation}
x_g=R_0\left(1-\frac{x}{2\,R_0}\right)^2 \, .
\label{eq:centre_mass_brut}
\end{equation}  
When the ball gets in contact with the ground, $x=0$ and $x_g= R_0$, the ball is not yet deformed and still a complete sphere. Both the center of mass and the center of the sphere are at the same location. For larger and larger indentation the two centres separate one from another, see Fig. \ref{fig:kinematics}.

\subsection{Dynamics of a non-dissipative inflated ball impact}
We explore, here, the dynamics of an inflated ball impact onto a rigid ground. All dissipation sources are neglected, no sound is emitted at the impact, no friction exists between the ground and the ball, gas compression during the impact is supposed adiabatic and reversible. Finally, the shell is assumed to have no thickness : no viscous dissipation occurs during the impact and no elastic energy is stored within the shell.
\subsubsection{Conservation of momentum equation}
According to Newton's mechanics, the product of the mass by the acceleration of the center of mass of an isolated object is equal to the external applied forces. The equation of conservation of momentum reads, 
\begin{equation}
m \ddot x_g = F_\mathrm{ext}
\label{eq:meca_Newton}
\end{equation}
In the chosen parametrisation, we compute the acceleration of the center of mass as a function of $x$ by double differentiation of Eq. (\ref{eq:centre_mass_brut}),
\begin{equation}
m \ddot x_g = -m \left(1-\frac{x}{2\,R_0}\right) \ddot x + m \frac{\dot{x}^2}{2\,R_0}\, .
\label{eq:center_mass}
\end{equation}
The right-hand-side of this equation is compounded of two terms that have a strong physical meaning. The first term corresponds to the variation of momentum due to the acceleration of the mass of the shell that is not is contact with the ground, $m \left(1-x/2\,R_0\right)$. The second term corresponds to the variation of momentum associated to the mass of ball that actually reaches the ground during an infinite time interval, $m \dot x/2\, R_0$, which velocity drops from $\dot x$ to zero. When the ball is not much squashed on the ground, $x\ll 2\,R_0$, the second term is much lower than the first one and the acceleration of the center of mass simply reduces to $\ddot x_g = -\ddot x$. 

In our framework, all energy is supposed stored in the gas (and nothing in the shell). One may then express the term $F_\mathrm{ext}$ of Eq. (\ref{eq:meca_Newton}). The force exerted by the ball onto the ground is the pressure inside the ball, $P$ times the surface of contact, $A$. The force is simply $F_\mathrm{ext}=P(x,t)\, A(x)$, where $P(x,t)$ is the pressure inside the ball that depends on the indentation $x$ and time $t$. It reads,
\begin{equation}
m \ddot x_g = P(x,t)\,A(x)\, .
\label{eq:meca_finale}
\end{equation}

\subsubsection{Gas compression}\label{sec:gas_compression}
To describe the motion of the ball, one has to compute the evolution of pressure as a function of time. We choose the perfect diatomic gas model, $P\,V=N\,k_B\,T$, where $V$ is the volume of the ball, $T$ is the temperature of the gas inside the ball, $N$ is the number of molecules trapped in the ball and $k_B$ is the Boltzmann constant. During the impact, the volume of the inflated ball is reduced to a fraction of its original volume. This involves an augmentation of the pressure and temperature of the gas inside the sphere. Since the ball vertical velocity is much lower than sound velocity ($\sim$ 343 m.s$^{-1}$ at 20$^\circ$C), it remains under the hypothesis of quasi-static compression. We also suppose that no heat is exchanged with the external medium, the entropy is constant. The compression of the gas is isentropic, and one may use the Laplace relation, 
\begin{equation}
P(x,t) V(x)^\gamma = P_0 V_0^\gamma\, ,\qquad \mathrm{or\, \, equivalently,} \qquad \frac{T}{T_0}=\left(\frac{V_0}{V}\right)^\gamma\, .
\label{eq:thermo}
\end{equation}

For a small compression of the ball, lets say $x=R_0/3$ (\textit{i.e.} $V_0/V = 1.08$), with an initial temperature of $20^\circ$C (absolute temperature of 293.15 K), the maximal temperature is $T = 53^\circ$C ($\gamma=7/5$ for a diatomic gas like $N_2$ and $O_2$ that together compose more than 98\% of air). 
If there is no loss of energy through the ball membrane, then all the energy stored in thermal energy is converted back in kinetic energy of the ball during decompression stage.

\subsubsection{Non dissipative impact}\label{sec:non_dissipative_impact}
We now have two coupled equations, Eq. (\ref{eq:meca_finale}) and (\ref{eq:thermo}), that describe the impact dynamics of an inflated ball. One may reduce this set of equation to obtain the master equation of impact of an inflated ball that only depends on $x(t)$ and its derivatives. 
%
Equations are set non dimensional, with time scale $t_0=\sqrt{m/2 R_0 P_0}$ and length scale $x_0=2R_0$,
\begin{equation}
-\left(1-x\right)\ddot x + \dot{x}^2 = \pi x \left(1-x\right) \left(\frac{1}{1+2 x^3-3 x^2}\right)^\gamma\, .
\label{eq:nonlinear_conservative}
\end{equation}
\begin{equation}
-\left(1-x\right)\ddot x + \dot{x}^2 = \pi x \left(1-x\right)^{1-2\gamma} \left(2x+1\right)^{-\gamma}\, .
\label{eq:nonlinear_conservative2}
\end{equation}

The equation of evolution of the indentation, $x$, during the impact thus depends on the mass of the shell and the thermodynamic coefficients of the gas. In the limit of small indentation, $x\ll 1$, this equation comes to, 
\begin{equation}
-\ddot x = \pi x \, .
\label{eq:simplified_eq}
\end{equation}
With the limit conditions $x(0)=0$ and $\dot x(0)= U_0\, t_0/x_0$, one finds, 
\begin{equation}
x(t) =\frac{U_0\, t_0}{\sqrt{\pi}\, x_0} \,\sin\left(\sqrt{\pi}\,t\right)\,.
\end{equation}

The dimensional solution is then, 
\begin{equation}
x(t)=U_0 \sqrt{\frac{m}{2 \pi R_0 P_0}}\, \sin\left(\sqrt{\frac{2\pi R_0P_0}{m}} t\right)\, .
\label{eq:dynamic_contact}
\end{equation}

The contact time is defined by the first time, $t>0$, at which $x(t)=0$. One finds, 
\begin{equation}
t_\mathrm{contact}= \sqrt{\frac{\pi m}{2R_0P_0}}\, ,
\label{eq:time_contact}
\end{equation} 
that does not depend on initial ball velocity for small indentation. We obviously have also $e=\dot{x}(t_\mathrm{contact})/\dot x(0)=1$, meaning that no energy has been lost during the impact. 

If we had solved the complete non-linear Eq. (\ref{eq:nonlinear_conservative}), the contact time would not have been constant. Indeed, for large indentation, the linear spring model of contact is not valid any more, it is replaced by a non-linear spring that get stiffer as indentation is increased (the elastic force is convex). Similarly to Hertz impact ($F\propto x^{3/2}$), the contact time is then expected to slightly decrease with impact velocity.

\subsection{How much elasticity is stored in the flexible shell?}
At the very beginning of this paper we did the assumption that gas was the main source of energy storage during the impact. Nevertheless, the shell is an elastic material which is deformed during the impact and thus may also store some energy. This is particularly true for hollow balls such as table-tennis balls, tennis balls or squash balls for which gas is expected to have little importance onto the bouncing behavior.

\citet{lazarus2012geometry} showed that for small deformations (point force indentation), a thin hollow spherical shell has a linear spring like behavior that depends on material properties, Young modulus, $E$, Poisson’s ratio, $\nu$ and geometrical characteristics, sphere radius, $R_0$, and shell thickness, $t_h$. The deformation energy of a flexible shell indented by $x$ reads, 
\begin{equation}
\mathcal{E}_\mathrm{shell}=\frac{1}{2}\frac{4 E\,t_h^2}{R_0\sqrt{3(1-\nu^2)}}\, x^2\, .
\label{eq:shell_energy}
\end{equation}
\citet{lazarus2012geometry} extend this result to the case of a pressurized shell with point force indentation,
\begin{equation}
\mathcal{E}_\mathrm{pressurized-shell}=\frac{1}{2}\frac{4E\,t_h^2}{R_0\sqrt{3(1-\nu^2)}}\underbrace{ \left(\frac{\pi}{2}\frac{\sqrt{\tau^2-1}}{\mathrm{arctanh}(\sqrt{\tau^2-1}/\tau)}\right)}_\mathrm{Pressure\,\, induced\,\, stiffness}\, x^2\, ,
\label{eq:shell_total_energy}
\end{equation}
where $\tau = P_0 \, R_0^2 \sqrt{3(1-\nu^2)}/ E t_h^2$ is the non dimensional pressure. We have here to comment on this expression. For $\tau<1$ (small inflation pressure), the pressure induced stiffness term is a real coefficient which is the ratio of two pure imaginary functions. This correction term is thus well defined and tends to 1 when $\tau$ tends to 0, we find back the expression of shell energy of Eq. (\ref{eq:shell_energy}). In the other limit, $\tau\rightarrow\infty$, Pressure induced stiffness is similar to $\pi \tau/ 2 \ln(2\tau) \sim \pi \tau/2$ since $\ln(\tau)$ does not vary much. In this limit one finds that the force is $2\pi P_0 R_0 x$ similarly to Eq. (\ref{eq:simplified_eq}).

Let's now compare the energy stored in the gas to the elastic energy stored in the shell. The energy stored in the gas, in the limit of non dissipative impact corresponds to the work of the gas force, r-h-s of Eq. (\ref{eq:nonlinear_conservative}) and expresses as, 
\begin{equation}
\mathcal{E}_\mathrm{gas}=\frac{1}{\gamma-1}\left(P\,V-P_0\,V_0\right)=\frac{P_0\,V_0}{\gamma-1}\left(\left(\frac{V}{V_0}\right)^{1-\gamma}-1\right),
\end{equation}

\begin{figure}

\begin{psfrags}
\psfrag{stored}[l][l]{Potential energy}\psfrag{dissipated}[l][l]{Dissipated energy}
\psfrag{s}[l][l]{shell stored energy}\psfrag{g}[l][l]{gaz stored energy}\psfrag{d}[l][l]{dissipated energy}
\psfrag{basketball}[c][c]{Basket}\psfrag{squash}[c][c]{Squash}\psfrag{table-tennis}[c][c]{Table tennis}\psfrag{volleyball}[c][c]{Volleyball}
\includegraphics[width=0.8\textwidth]{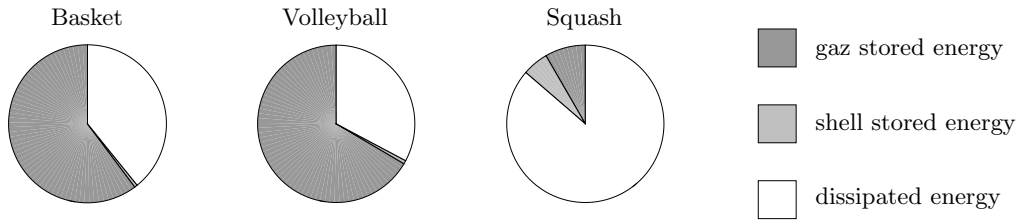}
\end{psfrags}
\caption{Diagram of energies involved during the impact of different sport balls (basket ball, volley ball and squash ball) at the maximal speed encountered in sport fields (\textit{cf.} Table \ref{table:ball_properties}). The total surface of the pie chart represents the incoming kinetic energy. This energy is separated in potential energy stored in the gas (gray) and in the shell (light gray) and dissipated energy that is lost during the impact (white).}\label{fig:diagramme}
\end{figure}

In the limit of small indentation, $x\ll 1$, it is simply, 
\begin{equation}
\mathcal{E}_\mathrm{gas}= 3 P_0\,V_0\left(\frac{x}{2R_0}\right)^2 \, ,
\end{equation}
which is consistent with the work of the approximated force $2\pi P R_0 x$.
We thus define the shell-to-gas elastic storage number, $S_t$, that reads, 
\begin{equation}
S_t=\frac{\mathcal{E}_\mathrm{shell}}{\mathcal{E}_\mathrm{gas}} = \frac{2}{\pi\sqrt{3(1-\nu^2)}}\frac{E}{P_0}\left(\frac{t_h}{R_0}\right)^2\, .
\end{equation}
Without surprise, $S_t$ scales like the inverse of the reduced pressure $1/\tau$, both compare elasticity to gas.
A small Storage number, $S_t\ll 1$ means that most of the energy is stored in the gas, whereas a large Storage number  $S_t\gg 1$ means that the shell elasticity is dominant for storing energy during an impact.
For a volley ball with mechanical characteristics, $E \sim 10^7\,\mathrm{Pa}$, $t_h\sim 2.1\, \mathrm{mm}$, $R_0\sim 10.5 \,\mathrm{cm}$, $\nu=0$ and $P_0=1.32\, 10^5\, \mathrm{Pa}$ (absolute pressure inside the ball), one computes, 
\begin{equation}
S_t= 0.01\ll 1\, .
\end{equation}
Most of the energy of a volley ball is stored in the gas. Storage numbers of different balls are gathered in Table \ref{tab:storage}. For large inflated balls used in basketball, football, handball and volleyball, $St$ is much smaller than one and the ball elasticity derives from the gas whereas for tennis, table-tennis and squash, the shell elasticity plays a major role. Finally, these conclusions can be summarized in the diagram of Fig. \ref{fig:diagramme} which shows the total energy involved in the impact of inflated sport balls. For each sport, the total energy is separated into the one stored in the gas and in the shell. Moreover, the diagram recalls the amount of energy which is dissipated in each situation.

\begin{table}
\begin{tabular}{l|ccccccc}
Sport&$E$ ($10^7\,$Pa) &$t_h$ (mm)&$R_0$ (cm)&$V_0$ ($10^{-3}\,$m$^3$)&$P_0$ ($10^5\,$Pa)&$S_t$\\
\hline
Basketball&1.0&2.5&12.15&7.5&1.55&1.0 10$^{-2}$\\
Tennis&0.7&4&3.18&0.137&1&0.37\\
Squash&0.1&8.2&2&0.033&1&0.61\\
Table-tennis&160&0.4&2&0.033&1&2.35\\
\end{tabular}
\caption{Storage numbers for usual balls. $\nu$ and $\gamma$ have been taken equal for all balls, $\nu=0$. For large inflated balls similar to basket ball (football, volleyball and handball), Storage number is always much smaller than 1. }
\label{tab:storage}
\end{table}

\section{Dissipation of energy during an impact}\label{sec:Dissipation}
Two quantities conveniently characterize the impact of a ball, the contact time, and the restitution coefficient. 
If the coefficient of restitution is 1, this means that the exit velocity is equal to the initial velocity and that no energy has been lost during the impact. This has been the case of study described previously in Section \ref{sec:non_dissipative}. In general, the restitution coefficient is always smaller than 1 for real systems. For a dissipative impact of a ball with $S_t\ll 1$, the force acting on the center of mass has a dissipative component $F_\mathrm{d}$. Eq. (\ref{eq:meca_Newton}) reads,
\begin{equation}
m \ddot x_g = P(x,t)\,A(x) + F_\mathrm{d}\, ,
\label{eq:meca_finale_dissipative}
\end{equation}
The dissipative force works along the trajectory and reduces the kinetic energy of the ball in the laboratory frame. In this section, we review all possible dissipation sources and estimate their approximative order of magnitude. All the modellings done hereafter use the spherical cap kinematics at the limit of non dissipative impact. We therefore use the Eq. (\ref{eq:dynamic_contact}) to describe the dynamics of the ball.

\begin{figure}
\begin{center}
\begin{psfrags}
\psfrag{x}[c][c]{$x$}\psfrag{2l}[c][c]{$2l$}\psfrag{gas}[c][c]{\textcolor{red}{gas}}\psfrag{R}[c][c]{$R_0$}\psfrag{dotl}[c][c]{$\dot{l}$}\psfrag{q}[c][c]{$t_h$}\psfrag{r1}[c][c]{$R_1$}\psfrag{p}[c][c]{\textcolor{red}{$(P,T)$}}
\psfrag{1}[c][c][0.8]{1}\psfrag{2}[c][c][0.8]{2}\psfrag{3}[c][c][0.8]{3}\psfrag{4}[c][c][0.8]{4}
\includegraphics[width=0.8\textwidth]{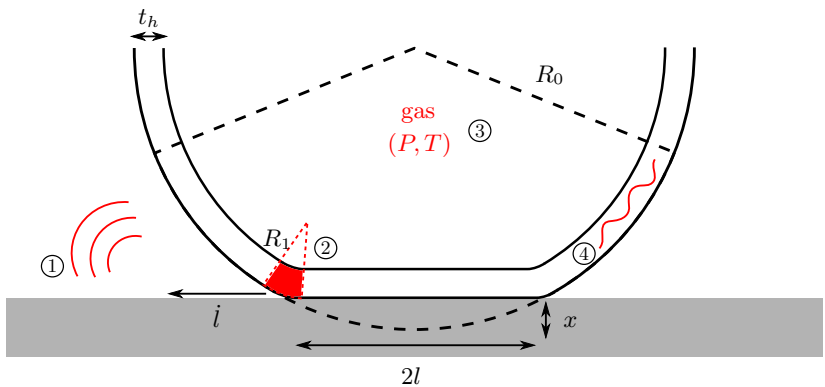}
\end{psfrags}
\caption{Zoom on the deformation of a ball during an impact, parametrization and sources of energy dissipation. (1) Sound wave emitted during the impact. (2) Viscous dissipation within the shell. (3) Dissipation in the gas during non-reversible thermodynamic process. (4) Transfer of kinetic energy into vibrations of the membrane.}\label{fig:dissipation}
\end{center}
\end{figure}

\subsection{Dissipation by sound}
When a ball impacts onto a solid, a sound wave is emitted and carries away some energy. We aim at estimating how much sound energy could be generated by an impact. 

The wave sound is emitted because of large lateral velocity during the impact. For a spherical ball impacting a flat surface with a constant velocity $\dot x =  U_0$, the lateral velocity of the ball is extremely large because of geometrical effects. Precisely, the velocity of the triple line ball-ground-air is given by,
\begin{equation}
\dot l = \frac{\mathrm{d}}{\mathrm{d}t}\sqrt{2R_0x-x^2}= \frac{(R_0-x)\dot x}{\sqrt{2R_0x-x^2}}
\label{eq:sound}
\end{equation}
This velocity diverges when $x\rightarrow 0$, $\dot l \rightarrow \infty$. As long as the lateral velocity, $\dot{l}$, remains larger than sound celerity, $c_s$, a shock-wave is emitted and energy is dissipated. 

Physically this shock wave corresponds to air particles moving faster than sound celerity. Under those conditions, air is compressed. The energy of the shock-wave is kinetic energy and compression energy. As a first approximation, we will suppose that the energy transferred from the ball to the gas corresponds to the kinetic energy of the gas. We have, 
\begin{equation}
\mathcal{E}_{d, \mathrm{shock}} = \int_0^{t_\mathrm{contact}/2} \rho_\mathrm{air} \frac{\mathrm{d}V}{\mathrm{d}x} \dot{l}^2 \dot{x}\mathrm{d}t\, .
\end{equation}
An upper bound for energy dissipation is then, 
\begin{equation}
\mathcal{E}_{d, \mathrm{shock}}\sim\frac{\sqrt{2\pi}}{3} \rho_\mathrm{air} U_0^3 R_0^{3/2}\sqrt{\frac{m}{P_0}}\, .
\end{equation}
For a basket ball, impacting at a velocity $U_0=16\,\mathrm{m/s}$, with an air density $\rho_\mathrm{air}=1.2 \,\mathrm{kg.m}^{-3}$, we have $\mathcal{E}_{d, \mathrm{shock}}\sim 0.34 \,\mathrm{J}$.

This energy is much less than the typical energy lost during an impact, $\sim 30.5\,\mathrm{J}$. Further measurements are needed to confirm that losses by sound radiated energy are negligible during the impact of an inflated ball. 

\subsection{Dissipation by ball vibrations}

Dissipation could occur in the shell through ball vibrations as suggested by \citet{cross2014impact}. We have here to separate the vibrations in two different origins: (i) the vibration of elastic shells (\textit{i.e.} the elasticity of the shell stores the energy, $S_t\gg 1$, \textit{e.g} tennis-table balls) and (ii) the vibration of inflated membranes (\textit{i.e.} similar to the model developed in the Section \ref{sec:non_dissipative} where all the elesticity is in the gas and all the mass in the membrane, $S_t\ll 1$, \textit{e.g} basket balls).

The case (i) is the one described by \citet{cross2014impact} for table-tennis balls. It corresponds to the vibrations of an elastic shell. Those vibrations have been  measured in impact force measurements for stiff balls but are not directly measurable optically. The amplitude of the vibrations is extremely small, about hundreds of microns. Nevertheless they are sufficient to dissipate energy, following this scheme, vibrations of amplitude $A$ occur over all the thickness of the ball, $t_h$, at a frequency $f$. $f$ is estimated to be the first mode of vibration of the elastic shell, $f=\sqrt{E}/\pi\,R_0 \sqrt{\rho}$, \citep{lamb1882vibrations}. For a tennis table ball we compute $f\sim 2\,10^4$ Hz, which is in the order of magnitude of free vibrations of the ball measured experimentally, $10^4$ Hz \citep{cross2014impact}. The volume of dissipation is the one of the ball solid material, $V_\mathrm{shell}= 4\pi R_0^2 t_h$. All the vibrating energy is supposed not be back-converted into kinetic energy and is totally lost. The total energy dissipated through the vibrations is then, 
\begin{equation}
\mathcal{E}_{d,\mathrm{vibrations}} \sim \frac{1}{2} m {A f}^2\, .
\end{equation}
For a table-tennis ball, with vibration amplitude $A=1 \,\mu$m, we compute $\mathcal{E}_\mathrm{vibrations}\sim 0.67\, \,\mathrm{J}$. This gives an order of magnitude of energy losses for a table-tennis ball in the same order of magnitude as in the experiment, see Table \ref{table:ball_properties}.

The case (ii) is the one of an inflated ball, where the shell can be considered as a membrane that contains all the mass of the ball. All the elasticity is in the gas. Change in shape of the membrane reduces the volume of the ball (the sphere maximizes volume for a given area) and increases pressure. According to this scenario, the interplay between gas elasticity and membrane inertia creates vibrations. The typical frequency of vibration is $f\sim\sqrt{R_0 P_0/m}$, see Section \ref{sec:non_dissipative_impact}. The vibration amplitude is the deviation from spherical shape, estimated from observations at 1 mm for a Basket ball. The dissipated energy then expresses similarly to the previous one,
\begin{equation}
\mathcal{E}_{d,\mathrm{vibrations}} \sim \frac{1}{2} m {A f}^2
\end{equation}

For a basket ball,  $A\sim 1\,\mathrm{mm}$, $f \sim 1.7\,10^2\,\mathrm{Hz}$, one obtains $\mathcal{E}_{d,\mathrm{vibrations}}=9.5\,\mathrm{J}$. Thus, the energy lost by vibrations is smaller than the one observed experimentally. 

\subsection{Dissipation by solid friction}
This section would strongly be considered for balls in oblique impact. However, for balls in normal impact as considered here, the friction force would be working in the orthogonal direction of the motion inducing no loss of energy in the normal direction.

At the microscopic scale, solid friction corresponds to the plastic deformation of the rough surface of the ground material. Here we consider that the solid material is infinitely stiff and smooth, no deformations are expected.
We conclude that no energy is dissipated because of solid friction, thus, 
\begin{equation}
\mathcal{E}_\mathrm{friction} =0
\end{equation} 

This result is consistent with the conclusions of \cite{pauchard1998contact} when no buckling of the contact surface of the ball is observed. 

\subsection{Dissipation in the shell}
Dissipations due to a rapid change of curvature of the shell during the impact are expected. The shell changes curvature from $C=1/R_0$ to $C=0$ in the flattened part . During this change of curvature, it transits by a large bending curvature $C$. The strongly deflected part, Fig. \ref{fig:dissipation} zone {\scriptsize \encircle{2}}, links the angle of the tangent of the circle, $\arcsin(l/R_0)$ to a null angle over a distance proportional to the bending-stretching length, $\sqrt{R_0\,t_h}$, along the triple line ball-solid-air \citep{lazarus2012geometry}. The curvature of this zone is then $C\propto \arcsin(l/R_0)/\sqrt{R_0\,t_h}$ and its deformation scales as $\varepsilon\propto C \sqrt{R_0\,t_h} = \arcsin(l/R_0)$. This zone has a volume $2\pi l\, t_h^{3/2} R_0^{1/2}$ and is deformed for a time equals to the passage time of the triple-line, $\dot{l}/l$. The rate of deformation, $\dot \varepsilon$, is then 
\begin{equation}
\dot\varepsilon \sim \varepsilon \frac{\dot l}{l}\, .
\end{equation}
In the limit of small indentations, $x\ll R_0$, we have $\dot\varepsilon \sim \dot l/R_0$. The instantaneous power dissipated by bending the membrane reads, 
\begin{equation}
\mathcal{P}_{d,\mathrm{shell}}\sim \eta {\dot{\varepsilon}}^2 (2\pi l) t_h^{3/2} R_0^{1/2}  = 2\pi \eta t_h^{3/2}\frac{l\dot{l}^2}{R_0^{3/2}}\, ,
\end{equation}
\noindent where $\eta$ is the viscosity of the shell material. By time integration, one finds the energy dissipated by bending the membrane, 
\begin{equation}
\mathcal{E}_{d,\mathrm{shell}}= \int_0^{t_\mathrm{contact}}\mathcal{P}_{d,\mathrm{shell}}\,\mathrm{d}t\, ,
\end{equation}
with Eqs. (\ref{eq:dynamic_contact}), (\ref{eq:time_contact}) and (\ref{eq:sound}), one obtains,  
\begin{equation}
\mathcal{E}_{d,\mathrm{shell}}\sim 24.6 \frac{ \eta \, t_h^{3/2} U_0^{3/2}\,{R_0}^{1/4}\,{P_0}^{1/4}}{m^{1/4}}\, .
\end{equation}

For a basketball impacting at $U_0=16$ m/s with a typical viscosity for synthetic rubber $\eta\sim 0.2\,\mathrm{Pa.s}$, the dissipation is in the order of magnitude of $0.52 \,\mathrm{J}$. The dissipation by deformation of the membrane is a good candidate for inflated sports balls. However, the viscosity of the material of the ball (synthetic rubber) is not well known and has to be measured for $\dot \varepsilon$ in the range 100 -- 1000 s$^{-1}$ with a viscoelastic rheometer. The presence of sews or different layers are thought to increase the dissipation.

\subsection{Dissipation in the gas}
 During the impact, air is warmed up because of compression. This creates a net temperature difference  between the air inside the ball and the membrane, typically 33$^\circ$C, see Section \ref{sec:gas_compression}. Heat transfer arises between the inner air in the ball (where energy is stored) and the membrane (sink). For the seek of simplicity, we consider hereafter that the temperature of the gas is uniform in all the volume.

An upper bound for thermal dissipation in dry air, following this scheme is, 
\begin{equation}
\mathcal{E}_{d,\mathrm{gas}}= 4\pi R_0^2\sqrt{D t_\mathrm{contact}} \Delta T\, C_p \rho  \, ,
\end{equation}
where D is the diffusion coefficient of temperature in the gas. With $\Delta T = 33 \,\mathrm{K}$, $D=20\,10^{-6}\, \mathrm{m}^2\mathrm{.s}^{-1}$, $t_\mathrm{contact}=10 \, \mathrm{ms}$, $C_p=1.0 \, 10^3 \mathrm{J.kg}^{-1}\mathrm{.K}^{-1}$ and $\rho= 1.2 \mathrm{kg.m}^{-3}$, one computes $\mathcal{E}_{d,gas}=3.28\,\mathrm{J}$. Almost no energy is lost by thermal transfer to the membrane, the adiabatic hypothesis is quite well verified.

The second proposed mechanism is a non-reversible process during the compression step. This would increase the entropy of the gas (viscous dissipation within the gas or non reversible transformation of one component like water vapour). To model irreversible generation of heat, one has to compute the flow of gas within the ball and then the work of the viscous force in the gas. The net effect would be to have a gas temperature larger after the impact than the initial gas temperature. This hypotheses can be easily tested by inflating the ball with different gases. 

The dissipation of energy by viscous friction within the gas is also a good candidate. For the ball with $S_t>1$, most energy is stored in the shell and less in the gas, it is relevant to think that dissipation occurs in the shell as demonstrated by \citet{cross2014impact}. However, for balls with small $S_t$, it seems that since most of the energy is stored in the gas, it is also in the gas where it is dissipated.

\begin{figure}
\begin{center}
\begin{psfrags}
\psfrag{stored}[l][l]{Potential energy}\psfrag{dissipated}[l][l]{Dissipated energy}
\psfrag{shell}[l][l][0.8]{shell}\psfrag{gaz}[l][l][0.8]{gaz}\psfrag{ad}[l][l][0.8]{sound}\psfrag{bd}[l][l][0.8]{vibrations}\psfrag{dd}[l][l][0.8]{shell}\psfrag{cd}[l][l][0.8]{thermal losses}\psfrag{ed}[l][l][0.8]{other}
\psfrag{basketball}[c][c]{Basket}\psfrag{squash}[c][c]{Squash}\psfrag{table-tennis}[c][c]{Table tennis}\psfrag{volleyball}[c][c]{Volleyball}
\includegraphics[width=0.9\textwidth]{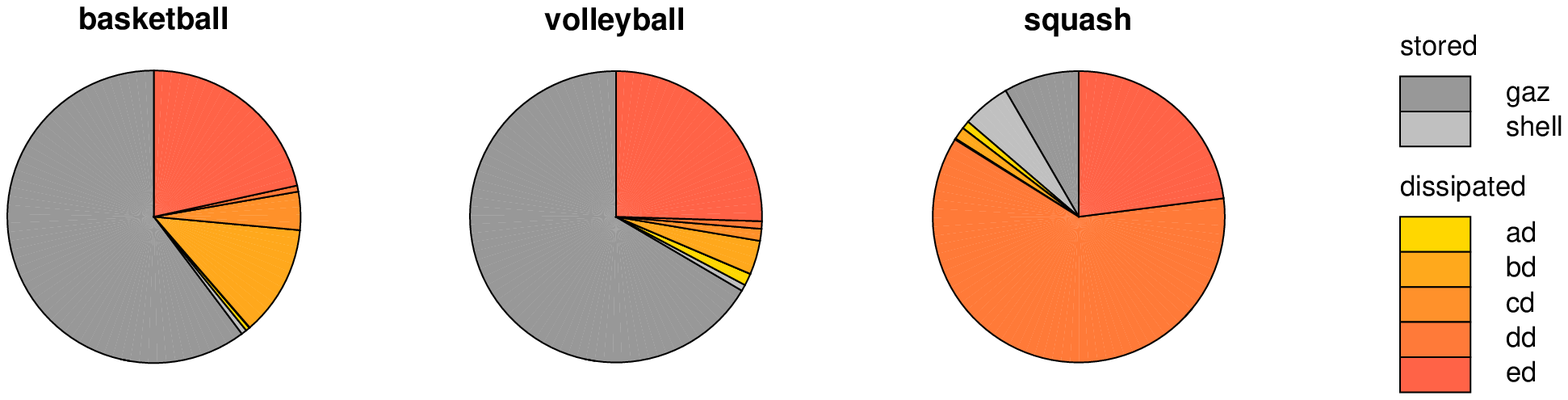}
\end{psfrags}
\caption{Diagram of energies involved during the impact of different sport balls (basket ball, volley ball and squash ball) at the maximal speed encountered in sport fields (\textit{cf.} Table \ref{table:ball_properties}). The total surface of the pie chart represents the incoming kinetic energy. This energy is separated in potential energy stored in the gas (gray) and in the shell (light gray) and dissipated energy that is lost during the impact (orange colors). The red couloured part represents the undetermined energy lost, either by gaz friction inside the ball or other sources of dissipation.}\label{fig:diagramme_complete}
\end{center}
\end{figure}

\section{Discussion}
All these effects may play a part into the dissipation of energy during an impact. The only model of dissipation of force presented until now, has been proposed by \citet{stronge2007oblique} and is called the \textit{momentum flux force}.
\subsection{Momentum flux force dissipation}
\citet{stronge2007oblique} proposes the moment flux force dissipation as a source of dissipation in the ball during an impact. This model states that the moment flux force of dissipation is equal to the second part of the acceleration of the center of mass, Eq. (\ref{eq:center_mass}), namely $F_d = -m \dot{x}^2/2\,R_0$. We have some comments about such a form of dissipative force,

(i) No microscopic explanation of transformation of energy unless suggestion that this force may account for ball vibrations is given.

(ii) The predictions of coefficient of restitution $e$ made by \citet{stronge2007oblique} for different impact velocities are lower than those actually measured even if they give good tendencies of evolution with impact velocity.

(iii) No extensive measurements of coefficient of restitution have been done varying inflation pressure and shell mechanical characteristics.

Extensive measurements of coefficient of restitution in different situations are thus needed to confirm or not this theory.

\subsection{Proposition of experiments}
To decide between \textit{momentum flux force dissipation} hypothesis and among other sources of energy dissipation mentioned above, we propose to make experiments on an idealized ball.

Systematic measurements of restitution coefficients have to be undertaken on an idealized ball by changing inflation pressure, $P_0$ and impact velocity $U_0$. The impact of the ball would have to be recorded at high speed to measure the impact characteristics. The measurements of contact time, $t_contact$, and restitution coefficient, $e$, would have to be done systematically to determine the relative importance of each of these depending on inflation pressure and impact velocity.

For vibrations of inflated balls, we propose to equip the ball with a accelerometer to measure microscopic vibrations of the shell. This accelerometer should have a frequency response higher than the first mode of vibration of the ball (10$^4$ Hz for a tennis table ball). Its pass band should also be in the range of the frequency of vibration of the ball. For large balls, direct optical measurements should be able to give the amplitude of vibration ($\sim 1$ mm). These vibrations would correspond to the \textit{momentum flux force dissipation}.

To measure the energy of the shock wave, we propose to record the sound emitted by the impact with a microphone. The interest is in estimating the sound energy by measuring frequency content, power and duration of the wave.

If dissipation occurs in the gas inside the ball, the equilibrium temperature inside the ball may change before and after the impact. To account for this change, we propose to record temperature evolution within the ball thanks to a fast-response thermocouple. By changing gas characteristics from water vapour saturated air to dry air (or pure diatomic gas like diazote, N$_2$) and finally monoatomic gas like helium, He, one should also be able to tell the importance of gas in the bounce mechanics.

For energy dissipation within the shell, It may tend to zero as the thickness of the shell is decreased. We propose to test the evolution of coefficient of restitution for similar ball (same mass, same radius, same material for the shell) with different shell thickness. This would necessitate to increase artificially the mass of the shell using small weights.


\section{Conclusion}
Non dissipative bounces of inflated balls have been modelled. It came out that the storage number, $S_t$, give the relative importance of the gas to the shell for storing elastic energy during the compression stage. For large $S_t$, most kinetic energy of the ball is converted in elastic energy of the shell (\textit{e.g.} tennis table ball). In the contrary for low $S_t$, most of the kinetic energy is stored in the gas (\textit{e.g.} basket ball, volley ball, foot ball, etc.) 

Before further comment, it is worth noting that this paper is a first attempt to give a comprehensive view of energy dissipation that occur during the impact of an inflated ball. One limitation is clear, we lack of experimental data. Nevertheless, we derived expressions for each sources of energy dissipation considered. We showed that some energy dissipation sources during an impact such as heat exchange and sound energy radiation are undoubtedly of weak importance for most of the sports balls. Concerning the other sources, conclusions are not trivial and depends on ball considered. For instance, tennis table ball, squash ball and basket ball have strongly different behaviours both in energy storage and in energy dissipation.

This paper has been initially motivated by a starting study of the authors about the optimal pressure of inflated balls. Rationalizing the damping of inflated balls has been a challenge definitively impossible to overcome just by theory and order of magnitude because of multi-physical effects which take place in this simple everyday life experiment. In conclusion, no consensus exists today about the mechanisms underlying restitution and dissipation during the impact of an inflated ball. We are convinced that an extensive experimental campaign is required to better understand this problem. Currently, no experimental proof definitely confirms the theoretical developments encountered in the literature about the behaviour of inflated balls during an impact. The approach proposed here, suggests that the bounce of an inflated ball is far to be as simple as a spring-mass-damper model. 



A better understanding of impact mechanics is essential for designing the next generation of sport balls and modelling their behaviour in sport fields. Ultimately, tackling the mechanics of inflated balls may lead to the development of sport balls with better bouncing properties. The mass of the ball, the inflation pressure, the properties of the inner gas and the mechanical characteristics of the shell would be chosen to have faster sports as well as more controllable balls while maintaining game safety. For instance, this would lead to the manufacture of extremely technical balls with constant bouncing characteristics whatever the altitude, field conditions, weather, etc. Moreover, the understanding of inflated balls mechanics will help to answer a crucial question occurring in sport such as: can you advantage a fast team against a technical team by playing onto the range of game prescribed inflation pressures?

Finally, investigating further this problem would also help to better understand why deflated balls advantage players for launching and receiving balls in American football. Until now, no quantification of the benefit gained by the Patriots during the deflategate has been computed.





\begin{thebibliography}{10}
\expandafter\ifx\csname natexlab\endcsname\relax\def\natexlab#1{#1}\fi
\def\au#1{#1} \def\ed#1{#1} \def\yr#1{#1}\def\at#1{#1}\def\jt#1{\textit{#1}}
  \def\bt#1{#1}\def\bvol#1{\textbf{#1}} \def\vol#1{#1} \def\pg#1{#1}
  \def\publ#1{#1}\def\arxiv#1{#1}\def\org#1{#1}\def\st#1{\textit{#1}}

\bibitem[Cross(1999)]{cross1999bounce}
{\sc \au{Cross, R.}} \yr{1999}  \at{The bounce of a ball}.  \jt{American
  Journal of Physics}  \bvol{67}~(3),  \pg{222--227}.

\bibitem[Cross(2014)]{cross2014impact}
{\sc \au{Cross, R.}} \yr{2014}  \at{Impact behavior of hollow balls}.
  \jt{American Journal of Physics}  \bvol{82}~(3),  \pg{189--195}.

\bibitem[Georgallas \& Landry(2015)]{georgallas2015coefficient}
{\sc \au{Georgallas, A.} \& \au{Landry, G.}} \yr{2015}  \at{The coefficient of
  restitution of pressurized balls: a mechanistic model}.  \jt{Canadian Journal
  of Physics}  \bvol{94}~(1),  \pg{42--46}.

\bibitem[Goodwill \& Haake(2001)]{goodwill2001spring}
{\sc \au{Goodwill, S.~R.} \& \au{Haake, S.~J.}} \yr{2001}  \at{Spring damper
  model of an impact between a tennis ball and racket}.  \jt{Proceedings of the
  Institution of Mechanical Engineers, part C: Journal of mechanical
  engineering science}  \bvol{215}~(11),  \pg{1331--1341}.

\bibitem[Hassett {\em et~al.\/}(2015)Hassett, Sullivan \&
  Veuger]{hassett2015football}
{\sc \au{Hassett, K.~A.}, \au{Sullivan, J.~W.} \& \au{Veuger, S.}} \yr{2015}
  \at{Football under pressure: Assessing malfeasance in deflategate}.
  \jt{Journal of Sports Analytics}  \bvol{1}~(2),  \pg{103--110}.

\bibitem[Lamb(1882)]{lamb1882vibrations}
{\sc \au{Lamb, H.}} \yr{1882}  \at{On the vibrations of a spherical shell}.
  \jt{Proceedings of the London Mathematical Society}  \bvol{1}~(1),
  \pg{50--56}.

\bibitem[Lazarus {\em et~al.\/}(2012)Lazarus, Florijn \&
  Reis]{lazarus2012geometry}
{\sc \au{Lazarus, A.}, \au{Florijn, H. C.~B.} \& \au{Reis, P.~M.}} \yr{2012}
  \at{Geometry-induced rigidity in nonspherical pressurized elastic shells}.
  \jt{Physical review letters}  \bvol{109}~(14),  \pg{144301}.

\bibitem[Lewis {\em et~al.\/}(2011)Lewis, Arnold \&
  Griffiths]{lewis2011dynamic}
{\sc \au{Lewis, G.~J.}, \au{Arnold, J.~C.} \& \au{Griffiths, I.~W.}} \yr{2011}
  \at{The dynamic behavior of squash balls}.  \jt{American Journal of Physics}
  \bvol{79}~(3),  \pg{291--296}.

\bibitem[Pauchard \& Rica(1998)]{pauchard1998contact}
{\sc \au{Pauchard, L.} \& \au{Rica, S.}} \yr{1998}  \at{Contact and compression
  of elastic spherical shells: the physics of a `ping-pong' ball}.
  \jt{Philosophical Magazine B}  \bvol{78}~(2),  \pg{225--233}.

\bibitem[Stronge \& Ashcroft(2007)]{stronge2007oblique}
{\sc \au{Stronge, W.~J.} \& \au{Ashcroft, A. D.~C.}} \yr{2007}  \at{Oblique
  impact of inflated balls at large deflections}.  \jt{International journal of
  impact engineering}  \bvol{34}~(6),  \pg{1003--1019}.

\end{thebibliography}
\end{document}